# Probing the Visualization Literacy of Vision Language Models: the Good, the Bad, and the Ugly

Lianghan Dong and Anamaria Crisan

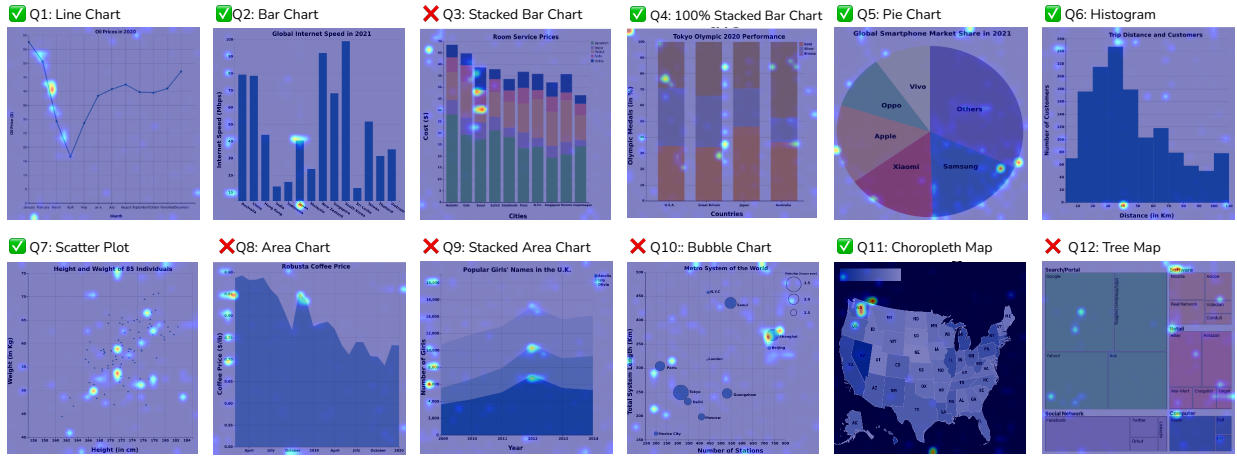

Fig. 1: Attention-guided Grad-CAM (AG-CAM) visualizations show the ChartGemma-3B [39] model's 'reasoning' over images and text questions from the Mini-VLAT [42]. Attention Saliency maps, highlighting important areas in red, are overlain on each chart image. ✓ and ✗ indicate correct vs incorrect responses, respectively. Each model layer and input token produces a unique result that can be used to interrogate *how* a Vision Language Model (VLM) produces its responses.

**Abstract**— Vision Language Models (VLMs) demonstrate promising chart comprehension capabilities. Yet, prior explorations of their visualization literacy have been limited to assessing their response correctness and fail to explore their internal reasoning. To address this gap, we adapted attention-guided class activation maps (AG-CAM) for VLMs, to visualize the influence and importance of input features (image and text) on model responses. Using this approach, we conducted an examination of four open-source (ChartGemma, Janus 1B and 7B, and LLaVA) and two closed-source (GPT-4o, Gemini) models comparing their performance and, for the open-source models, their AG-CAM results. Overall, we found that ChartGemma, a 3B parameter VLM fine-tuned for chart question-answering (QA), outperformed other open-source models and exhibited performance on par with significantly larger closed-source VLMs. We also found that VLMs exhibit spatial reasoning by accurately localizing key chart features, and semantic reasoning by associating visual elements with corresponding data values and query tokens. Our approach is the first to demonstrate the use of AG-CAM on early fusion VLM architectures, which are widely used, and for chart QA. We also show preliminary evidence that these results can align with human reasoning. Our promising open-source VLMs results pave the way for transparent and reproducible research in AI visualization literacy.
**Code and Supplemental Materials:** https://osf.io/fp3rg/?view_only=9b11aaec4ddf4656b205ebc53f4ef9db

**Index Terms**—Vision Language Models, Visualization Literacy, Explainability, Chart Question and Answering

---

## 1 INTRODUCTION

The increasing sophistication of large language models (LLMs) has expanded their use to diverse data analysis and visualization applications [58]. These include generating visualization code [38], developing visual analytic chatbots [56], providing support for visual design [28], data storytelling [49], and to some extent, visualization education [14]. However, LLMs alone cannot process, integrate, and generate insights from multiple data types, such as images and text. Consequently, their analytical capabilities to examine existing visualizations, for example, to answer questions or produce textual summaries, are constrained; using LLMs alone requires access to the underlying chart data, which is not always available. To overcome this limitation, new classes of multimodal models, notably vision language models (VLMs) have been developed [18, 59]. VLMs offer a distinct advantage by accepting and integrating image and text inputs, enabling them to generate textual analyses of visual content. This capacity allows for the interpretation of visualizations within media, such as magazine graphics, newspaper charts, or government reports, even in the absence of underlying datasets. However, the effectiveness of VLMs hinges on their ability to accurately synthesize and interpret both visual and textual information.

In this research, we interrogate the visualization literacy of VLMs. Visualization literacy is broadly defined as the ability to "read and understand" visual representations of data so as to draw "meaning from patterns, trends, and correlations" that they may contain [22, 23, 32]. Whether VLMs possess such abilities, or analogous reasoning skills, is an open question [26]. Prior research from the Natural Language Processing (NLP) and ML/AI research communities has prioritized exploring VLM's reasoning abilities with naturalistic images and scenes (e.g., containing cats and dogs, or people) [18, 59]. However, data visualizations present distinct challenges compared to naturalistic images, requiring the precise translation of visual encodings into quantitative data . The alignment between the chart content and a resultant response also requires understanding chart-specific and data-specific

• *Lianghan Dong and Anamaria Crisan [a7dong, ana.crisan]@uwaterloo.ca are with the University of Waterloo.*

semantics (e.g., "the upward trend on the scatter chart reflects growth in sales" [9]). While recent research has explored the visualization literacy of VLMs [5, 34, 37, 43], their use of proprietary (closed-source) models (e.g., GPT-4o, Gemini) limited their analysis to response accuracy results on standard tests (e.g., mini-VLAT [42], VLAT [31]).

*We probe the visualization literacy of VLMs by interrogating their internal reasoning when responding to questions about chart images.* Like prior work, we used questions from mini-VLAT [42] and VLAT [32] to examine the visualization literacy of VLMs. However, unlike prior work, which focused on evaluating the model's outputs, we used open-source models to explore what the models "see" when attempting to respond to a question. To do so, we developed a variation of Attention-Guided Gradient-weighted Class Activation Mapping (AG-CAM) [33] for Chart QA tasks to explore relationships and important features between image and text tokens when generating a response to a question. We use our approach to interrogate what different models "look at" when responding to mini-VLAT questions, what this reveals about their spatial and semantic reasoning capabilities, and how models compare to people's focus and reasoning. We summarize our findings along the following considerations:

- **The Good:** We show that small VLMs fine-tuned on chart QA tasks are nearly as performant as much larger closed-source models and with the added benefit of greater transparency. Using ChartGemma [39], we demonstrate the spatial and semantic abilities of VLMs and areas where they still struggle. We also show the AG-CAM can help us contrast VLM reasoning with that of people. Overall, we see a lot of potential for open-source models.

- **The Bad:** Even beyond overall performance, visualization literacy is not uniform across VLMs, like people they vary in their abilities. This issue may be more consequential for smaller models than larger ones and makes the choice of VLM important when embedding them into downstream applications.

- **The Ugly:** VLMs build their reasoning over layers and by learning relationships between text and image tokens. There is not always one single, simple, representative image that captures their entire reasoning processes. Moreover, interpreting the AG-CAM results can be more of an art than a science. However, this is not a fatal flaw, but rather opens exciting possibilities toward nuanced examinations, which we initiate here, into VLM capabilities.

Collectively our research makes the following contributions:

- We describe and release an adaptation of Attention Guided Grad-CAM (AG-CAM) for ChartQA on early-fusion VLMs (§4), including a companion application to explore its use.

- We conduct an exploration of visualization literacy of four open-source VLMs, using AG-CAM to examine their spatial and semantic reasoning abilities. We summarize VLM strengths and limitations and conduct a preliminary comparison to people.

- Discussions and future directions toward open-source VLM use and challenges and opportunities for exploring their 'reasoning'.

The rapid adoption of VLMs by researchers, practitioners, and corporations warrants closer scrutiny. Our research provides a mechanism and preliminary results for probing the capabilities of these models, setting up fruitful avenues for future work and experimentation.

## 2 RELATED WORK

We summarize pertinent prior research for assessing visualization literacy (in humans and for VLMs), chart question-answering, and finally AI model explainability.

### 2.1 Assessments of Visualization Literacy

Pioneering research into visualization literacy proposed various direct and indirect methods of assessments. Boy *et. al.* [8] proposed using item response theory to directly assess visualization literacy on a small range of charts – this work inspired VLAT and follow-on research. While Börner *et. al.* [10] explored the general public visualization literacy indirectly through a museum setting. More recent research has focused more on direct testing. A widely used approach for studying visualization literacy is the VLAT [31], a 53-item questionnaire that poses multiple questions using 12 chart types and relating to eight visualization tasks. A recent derivative, the mini-VLAT [42], demonstrates similar performance from a shorter assessment. Both assessments are widely for assessing visualization literacy in people. Nobre *et. al.* [41] uses VLAT to identify barriers to visualization literacy. Hedayati and Kay [22] used a modified 36-item version of VLAT to explore what university students learn in visualization classes. The VLAT test has also been shown to correlate with other cognitive characteristics (e.g., numeracy), that are generally essential for understanding and working with data. There also exist variations and extensions to VLAT. Critical Thinking Assessment for Literacy in Visualizations (CALVI) addresses the potential for readers to be misinformed by visualizations and draws inspiration from the VLAT. Saske *et. al.* [47] develop a Multidimensional Assessment of Visual Data Literacy (MAVIL) to assess visualization and data literacy along six dimensions, ranging from cognitive capabilities to aesthetics elements, including avenues for self-assessment of literacy skills. Taken together, all these tests roughly arrive at the same conclusion, which is that visualization literacy is a testable skill, varies in the human population, and should be considered when assessing data visualizations.

These same tests are now applied to assess vision language models (VLMs). Early work by Haehn *et. al* [20] explored CNN graphical perception, a complementary, but not identical to visualization literacy. More recent studies focus on VLMs. Hong *et. al.* [23] found that Gemini [19] and GPT-4o [1], using a modified VLAT, underperform human baselines, relying on internal knowledge. Conversely, Li *et. al.* [34] showed comparable or superior VLM performance (GPT-4o, Claude 3 Opus, Gemini) to humans baselines. Benedeck and Statsko [5] examining only GPT-4o, further confirms that VLMs excel at certain visualization tasks and struggle with others. Pandey and Ottley [43] used a unique prompting strategy to benchmark model performance (GPT-40, Claude, Gemini, and Llama) on both VLAT and CALVI to delineate the strengths and limitations of VLMs across visualization tasks. Finally, Lo and Qu [37], exploring explore that closed-source (GPT-4o, Gemini, Co-pilot) and open-source models (LLaVA-1.5 7B [35]), found evidence that VLMs possess sufficient visualization literacy to interpret misleading charts. However, because these studies prioritize closed-source models, they limit insight into VLM internal reasoning.

*Unlike prior work, our research attempts to probe what VLMs attend to and prioritize when responding to a prompt. In doing so, we provide another critical lens for examining the visualization literacy of VLMs.*

### 2.2 Chart Question and Answering

VLMs can be used for a variety of tasks, from generating captions, text-to-image search, retrieval, and question answering [24, 25]. We focus on Chart Question-Answering (QA) tasks; the VLAT is essentially a specific version of Chart QA. Islam *et. al.* [26] conducted a evaluation of three closed-source (GPT-4o, Gemini, and Claude) and several open-source models (Phi-3 [2]) against seven benchmark datasets of chart-specific tasks (e.g. QA, captioning, fact-checking, etc.). Their results demonstrate the VLMs do possess impressive capabilities for a variety of chart tasks, but also encounter common problems (e.g. hallucinations, sensitivity to prompts) and make errors. They also show that Phi-3 can be more performant than the closed-sourced models on zero-shot chart QA tasks. For some reason, prior research into visualization literacy has omitted open-source models. However, several exist that are fine-tuned specifically for chart QA and understanding, for example, ChartGemma [39], ChartLLama [21], and ChartAssistant [40]; of these ChartGemma is the most performant and we investigate it here. We think this is an oversight as performant open-source models are more transparent than their closed-source counterparts and enable the reproduction of experimental results because the weights are available. This is important given the rapid evolution of closed-source VLMs, which produced different results even in closely timed studies ( [5, 34, 37, 43]).

*Here, we continue to explore the capabilities of open and closed source models and ChartQA tasks. However, unlike prior which focuses just on the outputs, we also examine the model's internal reasoning.*

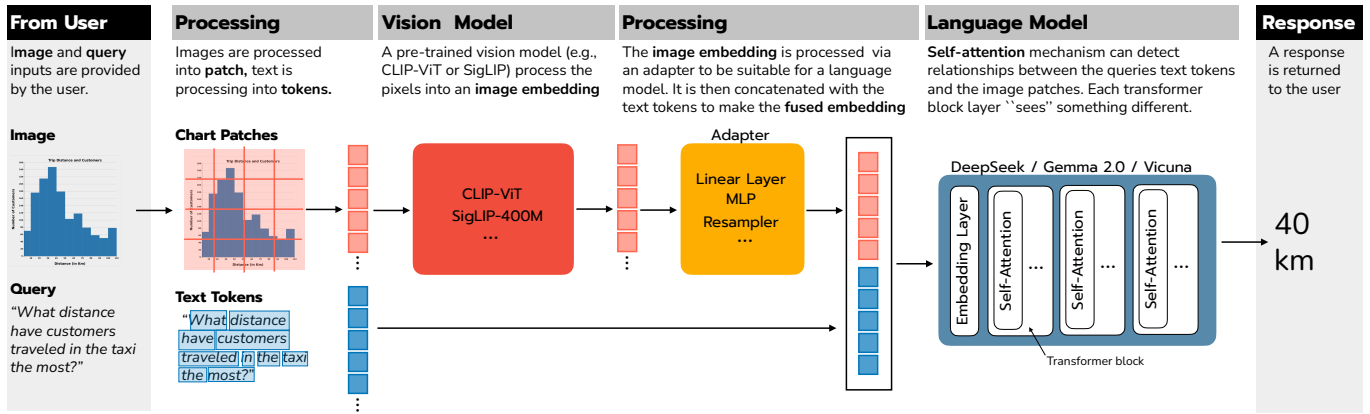

Fig. 2: Illustrative Overview of Vision Language Models. Here, we primarily explore the visualization literacy of early fusion (§3.1.1) architecture. We contribute an adaptation of attention-guided Grad-CAM for Chart QA to examine the internal reasoning of VLMs.

## 2.3 Visual Explanations for Vision and Language Models

Explainability in Vision-Language Models (VLMs) is particularly challenging due to their processing of heterogeneous inputs and the inherent complexities of both vision and language transformer components [30]. The ideas we explore here are drawn from prior research exploring class activation in CNNs and the visualization of attention mechanisms in transformer models. Gradient-weighted Class Activation Mapping (Grad-CAM) [48] is a widely used method of visually explaining important features, learned by CNNs. Grad-CAM generates a heatmap by leveraging gradients from the final convolutional layer, visually indicating which image regions contribute most to a target prediction (e.g., a cat or dog). It produces a saliency map that attributes individual pixels of the input image to the model's classification. Note, that there are several approaches, including gradient-based methods, that can produce a saliency map. Grad-CAM, while widely used, cannot work on the vision transformer (ViTs) models of VLMs, not only because of their different architecture but also because of how ViTs use patch-based processing for input images. While the ideas of this approach are still possible to use, they must be adapted for ViTs.

Visualizing attention is prevalent in transformer architectures and has been explored in prior visualization research [51, 54]. Self-attention works by learning weighted token relationships, which reveal relevant information within token sequences [50, 51]. To preserve semantic meaning across layers, attention aggregation requires careful handling [11]. Attention rollout visualizes information flow by recursively multiplying attention matrices, highlighting input regions influencing output, and is represented as an attention map [3]. Conversely, Layer-wise Relevance Propagation (LRP) propagates relevance scores backward; the idea was initially developed for CNNs and it was later adapted for ViTs [11, 52]. However, while self-attention indicates what a model attends to, it does not reveal the influence of individual input elements on the final response. Combining attention with gradients (e.g., [33, 61]) attempts to address this limitation, bringing together the best of both worlds. However, this combined approach has been primarily developed for and demonstrated in classification tasks and its use for more complex QA tasks requires further development.

*We adapted an attention-guided GradCAM approach to visualize the reasoning process of VLMs when responding to chart QA tasks. We show this reasoning process through attention-saliency maps overlain on chart images. To our knowledge, this is the first application of attention and gradient-based methods to probe visualization literacy.*

## 3 BACKGROUND: VISION LANGUAGE MODELS

Ahead of discussing our methodology, we first present a brief overview of VLMs and the specific architectures that are the focus of our work.

### 3.1 VLM classes and Architectures

Vision language models are a type of multi-modal model that can learn simultaneously from images and text [18]. There are different categories of such models. The first is multi-modal vision language models that take an image ($I$) and some query ($Q$) as input and produce a text response, for example, answering some question about a chart [24]. These models have three components: a vision transformer (e.g., CLIPs' ViT [44]), and adapter, and a separate text transformer (e.g., Vicuna [15]). In Figure 2 we show how these components work together and we describe this process in more detail in §3.1.1. Another category of models, which we do not explore, are text-to-image generation models, for example, DALL-E2 [45] or Imagen [46]. These models rely on diffusion models to synthesize images and often incorporate CLIP-based image and text alignment mechanisms to ensure a coherent output [57]. However, these models are not suitable for our investigations, because they do not produce text outputs, despite being trained on text. Their primary function is image generation rather than image understanding or reasoning.

### 3.1.1 Early vs Deep Fusion VLMs

Vision-language models are typically classified into two main types, *early fusion* and *deep fusion* models, based on their underlying learning mechanisms. While both types share a common architecture (see Figure 2), they diverge in how they integrate and jointly learn from text and images. Early fusion (EF) models concatenate the outputs from the vision component to the query tokens, which are then jointly processed by the language model component. Deep fusion (DF) models do not concatenate the vision and the query tokens but use cross-attention mechanisms to integrate the query and image embeddings into the different transformer block layers of the language model component. Early fusion models do not use cross-attention, but can still learn relationships between image and query tokens through self-attention mechanisms (we will describe the significance of this shortly). If the reader is unfamiliar with attention mechanisms, a pithy explanation is that it is a dynamic weighting system that allows the model to focus on the most informative features in a sequence of tokens. Self-attention operates on a single sequence, while cross-attention can operate on multiple sequences. Between the two approaches, EF-type models are the most widely used because of their architectural simplicity, computational efficiency, and the ease of using different pre-trained vision and language models without significant resource overhead. As they are the most common class of VLMs, we explore them here; comparing both the EF and DF models would be intractable in the scope of this work as both operate in different ways and would require different treatments.

### 3.1.2 Early Fusion VLM Pipeline

As a final piece, we summarize the early fusion VLM pipeline shown in Figure 2. To begin, a person can provide some query (e.g., "What distance have customers traveled in the taxi the most") and some input (e.g., a histogram image), and the VLM produces a textual response (e.g., "40 km"). In our research, these queries are drawn from the mini-VLAT and VLAT assessments. Both the query text and image pixels are processed in a manner according to their data types.

For images, each is re-sized to a common dimension and then split into non-overlapping patches. These patches have a standard size (e.g., [16 pixels x 16 pixels x 3 RGB layers]) that are individually flattened into a vector and then concatenated into a sequence of size $[S_I, f]$[1] where $S_I$ is the total number of patches and $f$ is the dimensions of the flattened patch vector. For images, the patches are processed by a pre-trained Vision Transformer (ViT) to produce an initial `image embedding`. The ViT used in VLMs comes from a simpler model, for example, CLIP [44] or some of its more recent variations (e.g., SigLIP [60]), which use a contrastive pre-training method that already learns some associated between text and images. This is important only for how embeddings are constructed; these ViTs, while suitable for classification, cannot generate text. The `image embedding` is the raw encoded representation of visual information of the image; it represents the image data in the ViT image latent space only and has not yet been connected with the query. The `image embedding` is further processed by an adaptor, whose role is to transform the embedding vector to match the language model's input space. We refer to this output as the `adapted image embedding`, it has size $[S_I, f']$.

Text is simply tokenized resulting in a $[S_Q, v]$ vector, where $S_Q$ is the total sequence length and $v$ is the vocabulary size of the tokenizer. The `adapted image embedding` is eventually concatenated with the text tokens of the query[2]. The concatenated image and text tokens produce the `fused input embedding`, which is processed by the language model. The self-attention layers of each transformer block can learn relationships between the text and image tokens, which we can directly observe and use to interrogate the model's internal reasoning.

## 4 METHODOLOGY

In this section, we describe how we adapted Attention-Guided Gradient-weighted Class Activation Mapping (Grad-CAM) from Leem and Seo [33] to visualize the reasoning processes of early fusion VLMs. The key differences between our approach and theirs are as follows. First, Leem and Seo [33] developed their method for classification tasks. We implement an extension, summarized in Algorithm 1, that makes it applicable to QA, which requires dealing with sequences of text inputs *and* outputs. Second, their method does not return results for each token $q \in Q$, whereas our approach does. Finally, our approach can return results for a single layer or any range of layers in the model, enabling greater flexibility to probe its internal reasoning.

### 4.1 Attention-Guided CAM (AG-CAM) for Chart QA

AG-CAM is an integration of two ideas, Grad-CAM, which quantifies the influence of individual elements of $I$ influenced the model's response, and self-attention, which tells us about the relationships between tokens in a sequence ($S$). The idea can be simply expressed as:

$$A_{ij} \cdot grad_{ij} \quad (1)$$

Each element $A_{ij}$ of the attention matrix A represents the amount of information flowing from a token $j$ in the preceding layer ($k$) to token $i$ in the subsequent layer ($k+1$), and where $i$ and $j$ are indices within the sequence of tokens $S = S_I \| S_Q$. Weighting attention by gradients gives higher priority to relationships between tokens that influence the model's output. This is visualized in a *attention-saliency (AS) map* (Figure 3); prior work [33] just refers to this as a heatmap, but, we use *AS map* to be more precise in the uniqueness of this output.

#### 4.1.1 Computing Attention and Gradients

We conduct a forward pass to compute the attention for each of the model's layers the model's layers (transformer blocks, $K$; where $1 \leq k < K$) and attention heads ($H$; where $1 \leq h < H$):

$$F_{h,q}^k = softmax(A_{h,q}^k) \quad (2)$$

Leem and Seo [33] refer to normalized attention scores (aka attention weights) as the feature map ($F$), however, their approach *only computes the feature map for the [CLS] token*, a special token that is intended

---
[1] For the moment, we are ignoring the batch dimension
[2] Concatenation is more nuanced than we show here, see [53] for details.

---

**Algorithm 1** Attention Guided CAM for QA

1: $out \leftarrow$ model(**inputs)
2: $logits \leftarrow out.logits$
3: $y \leftarrow \sum_{s=1}^{S} \max_{v} logits_s^v$
4: $y.backward()$
5: Get $F$, $grad$ in layers $[start, end]$, where $grad$ refers to $\frac{\partial y}{\partial F}$
6: **for** $k = start, ...end$ **do**
7: $\quad L_h^k \leftarrow F_h^k \odot ReLU(grad_h^k)$
8: $\quad L^k \leftarrow \sum_{h=1}^{H} L_h^k$
9: $\quad$ Get $L_q^k$ at $q$th prompt of $L^k$
10: **end for**
11: $L_q \leftarrow \sum_k^K L_q^k$
12: Select the image part of $L_q$
13: Normalize, reshape, overlay on original image

---

to serve as the ViTs final classification response. *We extend this idea to compute the feature map for each token $q \in Q$ allowing for more flexible and fine-grained resolution.*

The backward pass computes the model's gradients. Once again Leem and Seo [33] propagate gradients from a final classification result ($y^c$), however, in chart QA there is no classification result. Note that while the VLAT questions are multiple-choice and could technically have a classification, we do not limit ourselves to this. To address this limitation, we can treat the generation of the response as multiple classification tasks for predicting each word. So, for each element of $S$, it can be classified into anyone of the tokenizer's vocabulary ($v$). We can sum the output probabilities as follows:

$$y = \sum_{s=1}^{S} \max_{v} logits_v^s \quad (3)$$

Independent of our efforts, Zhang *et. al.* [61] proposed a similar approach; we discovered this only in the final stages of preparing our work. However, like Leem and Seo [33] their approach does not return a result for each $q \in Q$. We also note that our results show better localization overall and even across layers (Figure 6) relative to [61].

#### 4.1.2 Generating Attention Saliency Maps

Finally, we can compute the attention saliency (AS) map for a single layer ($L$) or a slice of layers, for each token $q$ as follows:

$$L_q = \sum_{k=start}^{end} \sum_{h=1}^{H} F_{h,q}^k \odot ReLU(\frac{\partial y}{\partial F_{h,q}^k}) \quad (4)$$

where $[start, end]$ denotes the layers we select, with $1 \leq start \leq end \leq K$. We normalize and reshape the results of $L_q$ and overlay it on the input image to explore the model's reasoning. We use a rainbow color map scheme to visualize $L_q$, where dark blue spots are not areas of interest and brighter red areas indicate higher importance. While rainbow color maps have some controversy [55], they are also the current norms for saliency maps that we opt to retain. An additional consideration from Leem and Seo [33] is to optionally normalize $F_{h,q}^k$ by sigmoid($G(\cdot)$) because it highlights more relevant pixels compared to softmax, which can hyperfixate. We explored both approaches (Figure 3) but opted to use primarily softmax when reporting our findings, which is more common and provided better localization for charts (i.e., a more focused response). We also experimented with different aggregation methods for $L_q$; namely using multiplication as is done in attention rollout [3], but found that summing across layers captured more interesting artifacts. As we indicate in §4.2, we make all of these options available in our companion application.

*Overall, our approach generalizes both prior methods [33, 61] to show more complex relationships between image and text tokens across layer slices of the model's reasoning.* We also demonstrate that our approach can be applied across several models, of different sizes, and with varying vision and language model components.

Table 1: Performance of VLMs on Mini-VLAT [42]. VLM responses vary across runs, in keeping with prior work [43] we show the average performance (% correct) across 10 runs. We use the performance baseline to add context to our exploration of the model reasoning.

| Model | #Params | Language Model | Vision Model | Mini-VLAT Questions | | | | | | | | | | | |
|---|---|---|---|---|---|---|---|---|---|---|---|---|---|---|---|
| | | | | Q1 | Q2 | Q3 | Q4 | Q5 | Q6 | Q7 | Q8 | Q9 | Q10 | Q11 | Q12 |
| ChartGemma [39] | 3B | Gemma-2B | SigLIP-400M [60] | 1.0 | 1.0 | 0.1 | 0.9 | 0.8 | 0.9 | 0.9 | 0.0 | 0.4 | 0.0 | 0.9 | 0.1 |
| LLavA-1.5 [35] | 7B | Vicuna-v1.5-7B [15] | CLIP-ViT [44] | 0.7 | 0.0 | 0.0 | 0.0 | 0.6 | 0.0 | 0.2 | 0.0 | 0.1 | 0.0 | 0.6 | 0.2 |
| Janus-Pro [13] | 7B | DeepSeek-7B [17] | SigLIP-400M [60] | 0.0 | 0.9 | 0.4 | 0.0 | 1.0 | 1.0 | 0.6 | 0.0 | 0.0 | 0.0 | 1.0 | 0.0 |
| Janus-Pro [13] | 1B | DeepSeek-1B [17] | SigLIP-400M [60] | 0.0 | 0.0 | 0.0 | 1.0 | 0.7 | 0.7 | 0.3 | 0.0 | 0.0 | 0.0 | 0.7 | 0.6 |
| Gemini-2.0-Flash [19] | ? | ? | ? | 1.0 | 1.0 | 0.4 | 0.0 | 1.0 | 1.0 | 1.0 | 0.0 | 1.0 | 1.0 | 1.0 | 1.0 |
| GPT-4o [1] | ? | ? | ? | 1.0 | 0.8 | 0.8 | 1.0 | 1.0 | 1.0 | 1.0 | 1.0 | 0.6 | 1.0 | 1.0 | 1.0 |

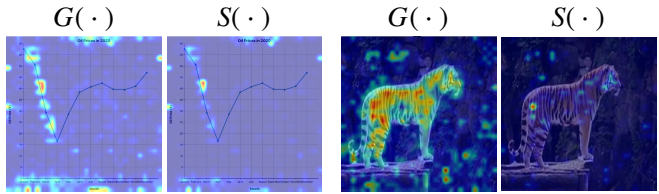

Fig. 3: Sigmoid ($G(\cdot)$) smoothing finds more correlations between pixels than softmax ($S(\cdot)$). While this may work well for natural scenes (e.g., the tiger), it introduces more noise for chart images. We implemented both, but primarily report on the softmax results. In this attention-saliency (AS) map, brighter colors up to red signify areas of more importance.

### 4.2 Implementation and Availability

We implemented our method in Python and released our code so that others could build our work. To view our results we also provide a lightweight Gradio application, a simple front-end framework that integrates natively with Python. Our application provides configurable controls to select different models, to set AG-CAM parameters, such as whether to apply $G(\cdot)$, to select layers to visualize, among other options. It has the mini-VLAT [42], VLAT [33], and an additional modified version of VLAT [41] available for use and it is also possible to upload images. Our application is available on Hugging Face [3]; to see a video of it in action please visit the supplemental materials. We do not claim this application as a contribution, it is just a companion to our work intended to increase its transparency.

## 5 PROBING THE VISUALIZATION LITERACY OF VLMS

We apply AG-CAM to compare what four different open-source VLMs "see" when attempting to answer mini-VLAT questions. We are interested not just in the variation of overall model responses, but also their internal processing, which we refer to as its "reasoning". We are motivated by two primary research questions:

**RQ1:** How do VLMs vary in their responses and reasoning?

**RQ2:** What kinds of reasoning capabilities do VLMs demonstrate and how does this compare to people?

We explore these questions along multiple dimensions. The first is the relationships between individual tokens in the query ($q \in Q$) and associated pixels in the image ($p \in I$). We did so by generating attention saliency (AS) maps and comparing them between models (§5.1.2). We focused more on tokens from $Q$ and $I$, over the response, because the model uses these to reason and generate an answer. The second dimension is between different types of charts, leveraging AS maps to examine the errors models make (§5.1.3). Finally, we look across the model's layers and how the AS maps compare to human annotations to make judgments of models' reasoning capabilities (§5.2). Throughout our investigations, we also made interesting and unexpected observations that we describe.

[3] https://huggingface.co/spaces/uw-insight-lab/Multimodal_Understanding/, running this space requires access to an Nvidia 1xL40S instance (or larger). Approximate costs are $1.80 USD/hour to operate (roughly $1,300 USD/month). Due to costs, we do not run this space continuously. However, it is possible to clone it and run it yourself.

### 5.1 RQ1: VLMs Fine-tuned on ChartQA Perform Best

Our first research question explores how different language models might reason about a chart image ($I$) and a user-provided query ($Q$) to produce a response. While we do present performance baselines (§5.1.1), we do not index our analysis on it. Instead, we use performance as a guide to explore *why* the model might have responded the way it did when probing its internal reasoning.

#### 5.1.1 Models and Baseline Performance

We primarily investigate four open-source VLMs. ChartGemma [39] is currently the most performant on images of data visualizations, having been specifically fine-tuned on Chart QA. As a basis of comparison, we also explore other models of similar size: Janus-Pro-1B [13], Janus-Pro-7B [13], and LLaVA-1.5-7B [35]. In Table 1 we show each model, its specific vision and language model components, and its results on the mini-VLAT questions. ***One difference from prior work is we require all models to generate the response, not choose among multiple-choice options.*** A numeric response is marked correct if it is close ($\pm 2$) to the actual answer (e.g., a response of 41 or 42 Mps, instead of 40). As VLM responses can vary, we report the average correctness from 10 model runs. We contrast the open-source model performance to the GPT-4o [1] and Gemini-2-Flash [19] closed-source models. While the specific details of closed-source architectures are not known at this time, we can be certain they are significantly larger than the open-source models we investigate. Of the smaller models, ChartGemma is the most performant, capable of rivaling even the closed-source models.

**Overfitting and Generalization.** The mini-VLAT [42] and VLAT [32], along with their responses, are publicly available. As such, it is reasonable to assume they may be part of VLM training datasets; if a valid assumption, the models can be overfit to mini-VLAT questions (e.g., it has memorized the answer). Prior research [23], attempted to mitigate these issues by creating new questions, while still using chart and task types of the original VLAT. However, we do not take the same approach here. By visualizing AS maps it would be evident if a model memorizes the answer but does not identify the appropriate elements of the chart to produce the results; the caveat being that this only works for open-source models (in our opinion, a clear benefit of their use). In fact, even if models are overfitting, our approach would allow us to interrogate if they are learning valid relationships between pixels and text tokens as part of that training process. As such, we are not as vulnerable to overfitting issues.

#### 5.1.2 Interpreting the Relationship between Text and Pixels

In Figure 4 we compare how the four open-sources models 'reason' and respond to the question "What is the average internet speed in Japan?" using the bar chart. The final column is a token that precedes the answer generation, which often, but not always, visualizes the answer. As we show in this figure, and later sections of this manuscript, reasoning is built over layers and tokens and one single image does not always capture the full result. However, this bar chart example is an exception. Using it as a reference, we summarize key observations on VLM performance, which can be explored via our companion application.

**All models can recognize text.** All models can make associations between a query token and text that appears in the image. For example, all models can identify 'speed' and 'Japan' in the chart axis and titles.

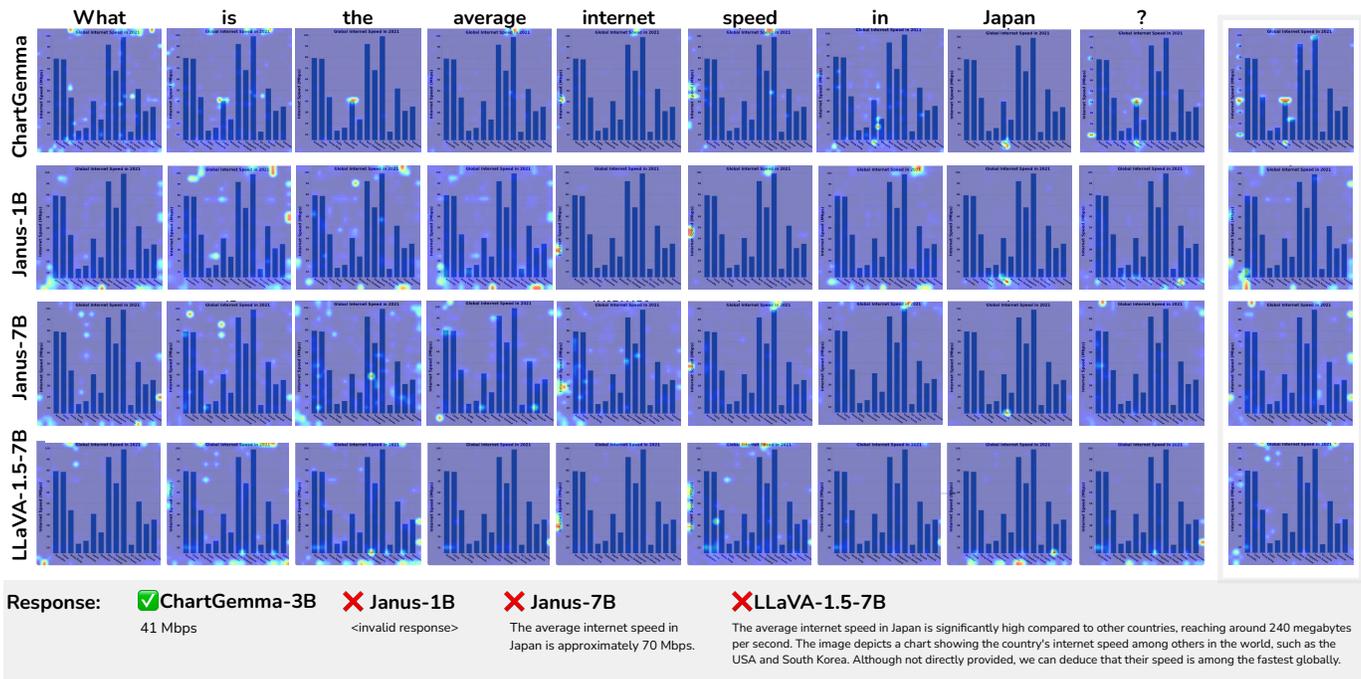

Fig. 4: Comparison of how different models "see" the chart image (*I*) in response to a common Query (*Q*) token. We show the attention saliency maps across four open source models (ChartGemma, Janus 1-B and 7-B, an LLaVA) across each word of the query *'What is the average internet speed in Japan?'*. The final column is an empty separator token, added by the models [6], which anchors the text generation of the output. Each model's final response is shown below, only ChartGemma is correct. While these models are of comparable size, and share common vision components, they behave differently; ChartGemma is the only model fine-tuned for ChartQA. *Note these are high-res images, zoom in to see details.*

Interestingly, all models also show stronger associations with 'Internet' in the chart axes over the title. Text can also be correctly read across the chart areas. For example, in Choropleth map example (Figures 1 and 5), ChartGemma makes associations between abbreviated and full state names (e.g., WA is Washington) and in the bubble, pie, and treemap charts it can identify labels (Figures 1, 6, and 8). VLMs are known to possess optical character recognition abilities [36], and so this result is not surprising, but, verifies the AS maps show expected associations.

**ChartGemma focuses more on visual encodings.** The Janus and LLaVA models do not show strong associations between $q \in Q$ and the visual encoding elements, whereas ChartGemma does. One reason for this is that ChartGemma is fine-tuned for ChartQA tasks, which not only has better overall performance (Table 1) but creates stronger associations between pixel and text tokens. Looking closely at the Janus and LLaVA models, there is some importance placed on elements of the encoding, particularly the horizontal tops of the bars which show a faint blue hue. However, compared to ChartGemma they do not ascribe importance to one bar (i.e., the one associated with Japan) over the others. This lack of focus on the encoding is reflected in the model's response as well. LLaVA's response appears drawn from its 'prior knowledge,' which refers to the information the model learned during pre-training from vast amounts of text and image data, and ignores the chart entirely. Both Janus models return the incorrect response; Janus-1B's invalid result is a non-sense computation. A final observation is that AS maps for ChartGemma appear more focused (e.g., show more specific areas of importance) compared to other models. One possible interpretation of these results is that when there are weaker associations between text and image tokens, not only does the model produce an invalid response but, the response may ignore the image. This means that the model is relying on its pre-existing understanding of general concepts and patterns, rather than accurately processing the specific visual data presented in the chart. Zhang *et. al.* [61] demonstrated something analogous to this when they conducted an experiment truncating image tokens, inducing the model to eventually indicate it could not respond. Furthering their approach could proactively detect when models rely more on their prior knowledge.

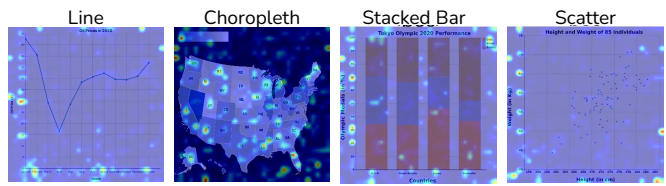

Fig. 5: The <BOS> (Beginning of Sentence) token is often added by the models as part of processing the inputs. It behaves as a starting point for a model to establish its context. In ChartGemma, the <BOS> token appears to focus on axes, text, and (faintly) aspects of the encodings (e.g., the shape of the line, points in the chart)

Finally, an interesting observation was the behavior of ChartGemma with the <BOS> (Beginning of Sentence) token, which is added by VLMs [6]. <BOS> signals the start of the sequence and can act as a contextual grounding for where the model will begin to build its understanding. Interestingly, in Figure 5 this appears to be the chart axis, which is a logical place to begin reasoning about a data visualization. We can observe the <BOS> token prioritizes different elements in different chart types, such as labels of all relevant text or encoding aspects of elements, such as the shapes of a line or preliminary trends.

> **Finding:** VLMs vary in how they balance textual and visual encoding information in charts. VLMs that are fined-tuned on chart QA tasks appear to balance these vision and text modalities better.

### 5.1.3 Performance Across Chart Types & Error Analysis

While we consider all open-source models, our analysis prioritizes ChartGemma as it makes the clearest associations with visual encodings. To conduct this analysis, we reviewed the mini-VLAT results for all models and made qualitative observations of their errors; future work can explore robust quantitative evaluations (§6). For ChartGemma we also explored the full 53-item VLAT from Nobre *et. al.* [41]. Across all models, we observed three primary types of errors relating to **Data**, **Encoding**, and **Reasoning**. These are not mutually exclusive, for example, data errors are related to issues with model reasoning.

[A]: **Data Errors** refer to inaccuracies in associating $Q$ with $I$, hindering the extraction of correct values for a valid response.

[A.1]: Look-up Errors concern instances where the model struggles to pinpoint the correct chart region. This was most apparent for Janus and LLaVA, which were capable of identifying text but often failed to make appropriate associations with visual encodings. By comparison, ChartGemma, even when incorrect, did identify the correct regions of the image; for example, in the stacked bar chart (Figure 1); Q: "*What is the cost of peanuts in Seoul?*"), ChartGemma seems to correctly identify the bar relating to Seoul and the part of the stack concerning peanuts. In the example of the bubble chart (Q:"*Which city's metro system has the largest number of stations?*"), it places the most importance on Shanghai (the correct answer), but, returns Beijing instead.

[A.2]: Extraction Errors pertains to problems retrieving accurate data values from the chart images. Once again, in the stacked *bar* chart example, the model appears to prioritize the correct parts of the image but ultimately does not produce the correct response. Another example is the stacked *area* chart, where the model accurately associates pixels representing the total number of girls with the names Amelia, Isla, and Olivia in 2012. It also gave higher importance to the values associated with Isla and Olivia, along with their corresponding positions on the x-axis. Still, the model generated an incorrect response, potentially due to difficulties in performing the ratio calculation.

[B]: **Encoding Errors** pertains to difficulties interpreting the shapes, trends, or patterns of visual encodings and relationships between marks.

[B.1]: Encoding Interpretation involves understanding individual data marks, such as associating a circle's size in a bubble chart or a wedge's shape in a pie chart to its data value (related also to Extraction Error [A.2]). This also includes interpreting pattern trends (e.g., decreasing line chart segments) or relationships between data marks (e.g., correlations in a scatter plot). We observed that ChartGemma (and occasionally Janus) performed effectively with simpler, more common chart types that are regularly used in media and public reports (e.g., line, bar, pie). They encountered greater difficulty with area encodings, complex spatial polygons (e.g., choropleth shape outlines), and charts involving relative baselines (e.g., stacked charts).

[B.2]: Hierarchical Relationships concern specific issues involving marks that are nested within each other. For example, in the tree map (Q:"True/False: eBay is nested in the Software category"), the model focuses on category names (e.g., Software, Retail, and Computer) but cannot resolve the hierarchical relationship with 'eBay'. While closed-source models performed better than ChartGemma here, it is uncertain whether this reflects a deeper understanding of these relationships or is the result of memorization (§5.1.1). However, neither mini-VLAT nor VLAT extensively assesses hierarchical relationships, such as those also found in phylogenetic trees or more complex composite encodings. This is a potent area for future work.

[C]: **Reasoning Errors** involve flaws in the model's ability to draw conclusions or make deductions based on the chart's elements.

[C.1]: Multi-step Reasoning concerns the number of analytic steps required to answer a question. For example, a bar chart retrieval task (Q:"*What is the average internet speed in Japan?*") can require just two steps: identifying the relevant part of the chart and extracting the result. Whereas, comparison or range tasks (e.g.,"*About how much did the price of a barrel of oil rise from April to August in 2020?*" in the line chart) require multiple retrievals and arithmetic operations (e.g., subtraction). Multi-step operations remain a challenge for VLMs, even with Chain-of-Thought prompting [29]. In Figure 9 we show that AS maps can help us interrogate this challenge and make progress.

[C.2]: Prompt Sensitivity concerns how the model's responses and reasoning may change because of how the model is prompted. To explore these types of errors we made modifications to the mini-VLAT questions in an *ad hoc* manner. We show in §5.2.3, Figure 7 and Figure 9, that the model's reasoning changes and impacts its responses.

> **Finding:** We observed three primary types of errors, **Data**, **Encoding**, and **Reasoning**, that can be interrogated with AS maps.

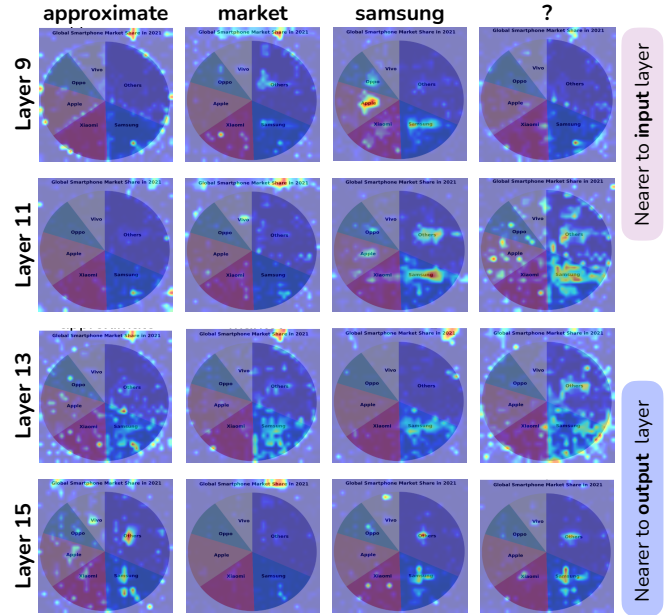

Fig. 6: The evaluation of reasoning over model layers. We use sigmoid normalization ($G(\cdot)$, §4.1.2) to show a fuller range of the model's focus.

## 5.2 RQ2: VLMs Reasoning Exhibit Higher-Order Reasoning

We now explore how VLMs construct reasoning over multiple layers (§5.2.1), which is a precursor for examining their spatial (§5.2.2) and semantic reasoning (§5.2.3). Here, we focus exclusively on ChartGemma and consider the 53-item version of VLAT reported by Nobre *et. al.* [41], which also includes annotations made by people that we compare to with AS maps (§5.2.4).

### 5.2.1 Building Reasoning Over Layers

Up to this point, we have shown results from one or more tokens, but only from a single layer. In Figure 6, we show how VLMs construct their reasoning over multiple layers (specifically, 9, 11, 13, and 15) from the query "What is the approximate global smartphone market share of Samsung ?"; we do not show more layers simply in the interest of space, but these can be explored in our companion application (§4.2). Each layer in Figure 6 is visualized individually without the contributions of prior layers (i.e., $start = end$, §4.1.2). Like prior work [61], we find that earlier layers in the model demonstrate more activity toward spatial understanding of visual encodings, while later layers contribute less. Individual layers also prioritize different information and do not just extend prior layers. For example, Layer 9 in the token 'approximate' highlights the importance of the pie's diameter (the other tokens do as well, but less so), while in Layer 11 this focus is absent. Instead, the model appears to prioritize text more. By Layer 13, the model seems to have localized the Samsung wedge. We generally observed that, across all chart types, Layer 13 in ChartGemma shows the most concrete examples of spatial reasoning that also relates to the model's final answer. Another relevant observation is the relationships between text and image tokens. For example, why does 'approximate' appear to emphasize the pie's diameter and '?' emphasize the diameter *and* the Samsung wedge? There isn't a clear answer. For terms like 'approximate' or 'average', which imply some mathematical operation, we have made a few observations that models, even Janus and LLaVA, attempt to 'look' at the broad spatial structure of the encoding. However, these observations were too few and qualitative to make more definitive conclusions. Lastly, interpreting precise token contributions can be complex because attention mechanisms dynamically assign weights that allow the model to focus on the most relevant information. The '?' token, which appears toward the end of the query, potentially receives more aggregated information from other tokens, leading to a contextually richer visualization of the response within a layer.

### 5.2.2 VLMs exhibit Spatial Reasoning

*We define spatial reasoning with respect to data visualizations to concern understanding the bounds of an encoding mark and its channels (e.g., color, shape, size) and relationships between multiple marks.* From the previous examples, it is clear that VLMs so possess some, but not perfect (§5.1.3), abilities to reason spatially. The most salient example of this was the ability of the model to clearly trace the lines (radii) and arc of the Samsung wedge in Figure 6. Earlier research [34] suggested that VLMs (specifically Claude, GPT-4o, and Gemini) might not understand the angled shapes of pie charts, as observed through post hoc experiments; our results contradict their observation. ChartGemma also demonstrated the ability to "see" a variety of spatial patterns, for example, the downward trend of the line chart or the linear relationship between points in the scatter chart. In the Choropleth map, it also appeared to extract values from the color scale by attending to the legend. Prior research on VLMs could not demonstrate this spatial reasoning because the researchers did not have access to closed-source models' weights. Knowing now the capabilities of open-source VLMs can rival those of close-sourced models, it is possible to transparently explore the limitations of spatial reasoning in ever more complex and bespoke chart types that the mini-VLAT and VLAT tests do not consider.

> **Finding:** VLMs trained on Chart QA tasks possess spatial reasoning capabilities. That is, they can recognize marks and channels, with limitations, and the relationships between marks.

### 5.2.3 VLMs Integrate Spatial and Semantic Reasoning

*We define semantic reasoning with data visualizations as the ability to understand and infer relationships between elements of Q and I.* We have already shown (Figure 4 and 6), that models, but ChartGemma especially, can identify pixels in the image that correspond to words in the query. In Figure 1 there are more complex demonstrations of semantic and spatial understanding as well. In the scatter plot, there does not exist any explicit line that summarizes the relationship between the two axes (height and weight). It was fascinating to observe ChartGemma appear to draw such a line and then correctly interpret its direction (upward) and that this represented a positive correlation. Here, we explore the relationship between semantics and spatial reasoning capabilities a little further. In Figure 9 we show examples of queries, taken from the mini-VLAT, that ask whether some trend is increasing or decreasing. We can observe that across three chart types (Line, Area, and Stacked area) the model can make associations between 'increasing' and the parts of the encoding that show an increasing trend and vice-versa for 'decreasing'. As we have discussed previously (§5.1.3) this can vary by encoding type, with area charts being some of the more challenging, but the model nonetheless finds increasing/decreasing segments across all charts. Retaining all other words in the query, but replacing increasing/decreasing with rising/falling, respectively, we can see a change in the model's reasoning. For 'rising' the model continues to find the upward trending regions of the charts (the area chart excepted), while for 'falling' the model appears to no longer make an association with the downward trending regions. Bromely and Setlur [9] showed people ascribe diverse semantics to line chart trends, their results are pertinent in light of the VLM's sensitivity to wording.

> **Finding:** VLMs fine-tuned on chart QA tasks demonstrate semantic reasoning by identifying relationships between key elements of Q and I and, with some limitations, making appropriate inferences by extracting and appropriately processing relevant data.

### 5.2.4 VLM Reasoning Aligns with People

Finally, we explore the question of whether there is any alignment between what AS maps visualize and the reasoning of people. To do so we gathered examples from a previously published study by Nobre *et. al.* [41]. In their study, they collected free-form sketches and open-ended text responses that captured the participants' thought (reasoning) process when responding to a modified set of VLAT questions. We compare the AS maps we generate to participants' sketch annotations, once again summarizing key observations.

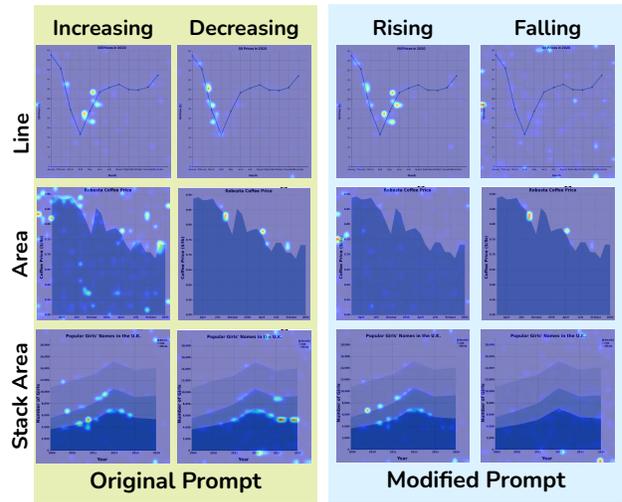

Fig. 7: VLMs are sensitive to the language of prompts, not just in their responses, but internal reasoning. Here, we change the increasing/decreasing terms to rising/falling, which have similar meanings. Notably, the VLM exhibits more difficulty associating 'falling' specific areas of the chart, but shows a stronger association with 'decreasing'.

There are some challenges in directly comparing AS maps to these annotations. First, as we have already discussed, model reasoning evolves over layers and tokens, yet, to compare to human annotations, we had to select a representative AS map to compare against. Second, as we show in Figure 8 humans also have different annotation strategies and can also focus on the wrong things. It is worth noting the reference human, like the reference VLM, matters as the accuracy range reported in [41] on tasks varies from 0.47 to 0.94. As such, it is not straightforward to treat these as ground-truth annotations to identify the mutual interest of the model and people [7]; our entire paper could be devoted to this topic alone. From our observations, the AS maps from ChartGemma do have reasonable alignment with regions of the chart that people also focus on. When people responded correctly (regardless of how the model responded) we noted two divergent strategies. First, people use the legend more consistently, while this is variable among VLMs. Sometimes, the AS maps show some evidence that the VLM attends to the legend (e.g., Figure 8-BUCQ3), but in other examples (e.g., the stacked area chart) it does not. *People appeared to attend to the legend more consistently than VLMs.* Second, for area encodings, which VLMs continue to struggle with, *people are better at tracing the shape of a mark* (e.g., tracing the outline of an area in the tree map).

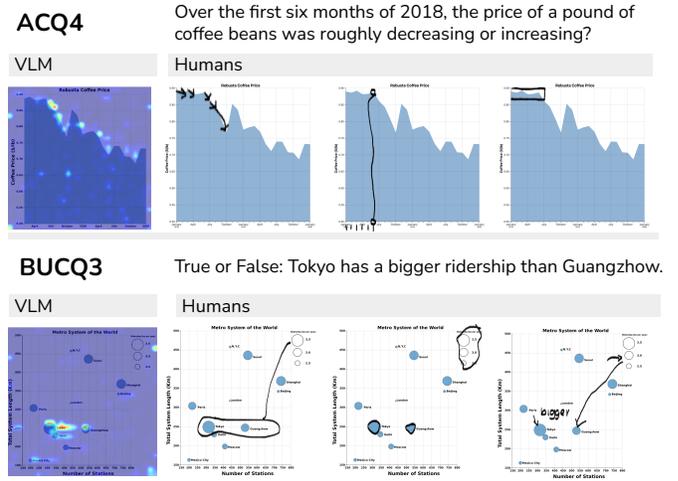

Fig. 8: Comparing our results with human annotations from [41]

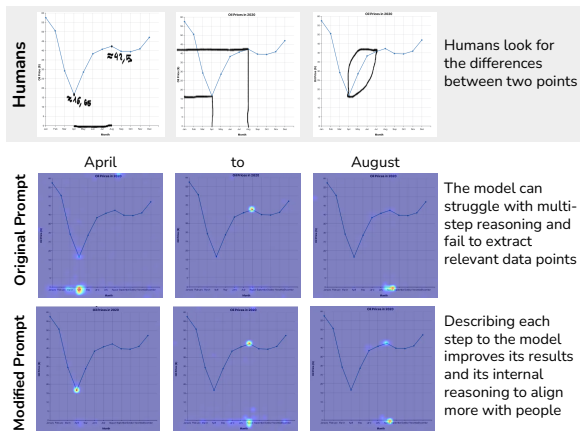

Fig. 9: Adding step-by-step instructions impacts reasoning & responses

We can explore the synergy between people and models further. In Figure 9 we show that humans distinctly annotate the two points necessary to answer the question. We could translate their interactions into steps as instructions to the model. Experimenting with this approach, we modified the original question prompt to provide step-by-step instructions ("*About how much did the price of a barrel of oil rise from April to August in 2020? Steps: First, extract the price in April. Then, extract the value of August. Finally, subtract and get results.* "). As a result, not only does the model put more importance on the April *and* August data points, aligning more with people, but it returns the correct response. Extending this experiment further can lead to novel avenues for model tuning with human guidance.

**Finding:** Our preliminary findings suggest VLMs and people can identify common regions of importance when reasoning with charts.

## 6 DISCUSSION

Visualization literacy requires reasoning over visual encodings to identify key elements (e.g., points, trends, patterns) and draw meaningful conclusions when answering questions. Prior research that explores these capabilities in VLMs only interrogates the model's behaviors and overall performance, limiting our understanding of their internal reasoning. We extended AG-CAM for ChartQA tasks to probe the visualization literacy of VLMs more deeply. We demonstrate that this approach reveals that VLMs possess spatial and semantic reasoning abilities that even appear to align with aspects of human reasoning. However, VLMs are not perfect and can struggle with complex encoding shapes (namely areas) and questions requiring multi-step reasoning. These limitations can be improved with further model tuning, including leveraging guidance from people. We now reflect on our findings.

### 6.1 The Good, the Bad, and the Ugly

Despite generally optimistic results concerning VLMs' visualization literacy, there are still areas for improvement, as reflected in our title.

**The Good: the Power of Small Open-Source Models.** A key difference between our work and prior research into visualization literacy is that we prioritized open-source models. This was not difficult as ChartGemma was nearly as performant as closed source counterparts, with the added benefit of being more transparent and, given the availability of weights, capable of generating reproducible results. ChartGemma is also small enough to be run on a performant gaming laptop. Overall, the architecture of early fusion models, which include ChartGemma, reduces their computational overhead relative to their deep fusion counterparts, making it easier for visualization researchers to experiment and extend its capabilities or propose alternative models. Moreover, we may not need to prioritize large general-purpose LLMs or VLMs, but could create specialized small models for visualization tasks. Given inequities in AI research [16], it is welcome news that it is not necessary to have a fistful of dollars to make progress. Beyond this, if we have more transparency toward the model's training data, we do not need to conduct overfitting arms races to define new VLAT questions [23].

**The Bad: Visualization Literacy is not Uniform Across VLMs.** In our research, and in keeping with prior work [5, 23, 34, 43], we consider visualization literacy to be a broad property that applies across all VLMs. However, VLMs, like people, vary in their abilities. For this reason, it is not precise to consider VLMs as generally having visualization literacy but that specific instantiations, even with the same general architecture, can be more capable than others. In our research, we found VLMs specifically finetuned on ChartQA to be more performant than those that are not and are of equivalent size. It is possible that with larger model sizes, the differences we observed between the smaller open-source models are diminished. However, given the benefits of small models (e.g., lower resource requirements, ability to embed on devices) [12], the specific choice of model will likely remain an important consideration in downstream applications. At least until some innovation reduces our reliance on scaling laws [27]. The good news is that our results suggest that producing specialized models for visualization tasks is possible and viable.

**The Ugly: the Murky Middle and Subjectivity.** When models' internal reasoning clearly aligns, or doesn't, with their responses, for example as in Figure 4, it can be straightforward to interpret AS maps. However, as others have pointed out, the interpretation of saliency maps is more a matter of art than science [4]. Moreover, considering that models create their reasoning across layers and tokens, it can be difficult to point to one singular representation of their reasoning. However, we believe this opens up exciting and potent avenues for future work. AS maps are another lens that can be used to probe the capabilities of these models, along with performance and other new approaches. For example, we can ask models to also explain themselves as Nobre *et. al.* [41] prompted people to do. Taking multiple lenses sourced from the models' behavior (e.g., explaining itself, accuracy) with its internal reasoning (e.g., AS maps) will help us create a more complete picture.

Complicating matters further is that, even in AI research writ large, defining and especially quantifying 'reasoning' remains an open problem. We provide specific examples of spatial and semantic reasoning, but our results are subject to the limitations of current understanding in the field of AI more broadly. Thus, future work examining the reasoning of both people and AI models requires a measured consideration that attends to the idiosyncrasies of both. Even when AIs and people prioritize common information, it should not be necessary for VLMs to entirely replicate human processes, but, rather be guided by them.

### 6.2 Limitations and Future Work

While this study offers valuable insights into VLM visualization literacy through the application of AG-CAM, it is important to acknowledge several avenues for future exploration. First, while our interpretation of attention saliency maps revealed compelling alignments between model responses and internal reasoning, particularly in fine-tuned models, we report primarily qualitative observations. As our goal was to probe and explore VLMs, we felt this was appropriate. Moreover, the challenge of quantifying reasoning is too broad to resolve in a single manuscript. However, throughout our work, we provided suggestions for future directions stemming from our observations, including experiments that could yield more quantitative insights. Second, the VLAT test suite, while useful for assessing baseline visualization literacy, may not fully capture the nuances of real-world chart understanding. Future research could incorporate more complex, real-world chart scenarios and evaluation metrics, leading to more ecologically valid results.

## 7 CONCLUSION

We examine the visualization literacy of VLMs using an adaptation of AG-CAM that we developed. Our results show the variations among VLM internal reasoning, with models fine-tuned on chart question-answering exhibiting the strongest overall performance. We also show that VLM reasoning can align with human reasoning, opening avenues for leveraging human guidance. Collectively, our findings demonstrate the value and viability of interrogating model reasoning, particularly by using small, open-source models. As VLMs continue to be adopted for real-world applications, our approach provides a valuable tool for the transparent and effective investigation of their visualization literacy.

## ACKNOWLEDGMENTS

We wish to thank the members of UW Insight Lab, F. Feng, X. Yu, and V. Bector for their feedback. We also wish to thank F. Shi for the discussion and suggestions. L. Dong is supported by a Cheriton School of Computer Science Undergraduate Research Fellowship.